# A Human Computer Interaction Solution for Radiology Reporting: Evaluation of the Factors of Variation

Authors: Tejaswini Ganapathi, David Vining, Roland Bassett, Naveen Garg and Mia Markey

## Abstract

**Purpose:**
The purpose of this research is to evaluate human and technical factors required to create a human-computer interface (HCI) for a radiology structured reporting process that incorporates eye-gaze and speech signals. Gaze and speech signals generated by radiologists during simulated image interpretation and dictation tasks were analyzed to describe the a) variation of gaze – speech temporal relationships in a radiology workflow environment, and b) variation in eye movements for a particular image interpretation task among different radiologists. Knowledge of these factors provides information regarding the complexity of the image interpretation and dictation task which can be used to design a HCI for future use in diagnostic radiology. Our goal is to use these data to create an HCI to automate the generation of a particular type of structured radiology report that captures key image findings and a radiologist's subsequent spoken descriptions of those findings to create a multimedia report [1].

**Materials and Methods:**
This study received Institutional Review Board approval. The radiologic images used in this study were de-identified to comply with HIPAA requirements. Five radiologists performed image interpretation and dictation tasks on 30 images from 3 modalities (10 images per modality), during which their eye gaze and speech signals were recorded using a Tobii eye-tracking device (Tobii T60 XL, Tobii Technology, Danderyd, Sweden). The dictation tasks were designed to be consistent across each modality and with each radiologist identifying two distinct targeted findings (e.g., identify and describe a lung mass on a chest x-ray followed by commenting on the heart size). Statistical modeling was performed to evaluate the variation of the eye gaze – speech temporal relationship on certain human and technical factors, such as the radiologist, image content, modality and resolution, and target order.

**Results:**
Our results indicate that there are substantial differences in the gaze-speech relationship among radiologists (p<0.001). There is no statistically significant evidence that the gaze-speech temporal relationship depends on the particular image modality (p=0.909) or target order (p=0.142). Analysis of gaze paths suggests that the search path variance among radiologists is significant.

**Conclusions:**
Our data indicate that a) gaze-speech relationships and b) scan paths vary significantly among radiologists, thus suggesting that a gaze-speech system for automating the capture of data for structured reporting processes may need to be customized for each user. The image resolution and layout, image content, and order of targets during an image interpretation session are irrelevant factors to consider when designing an HCI. Our findings can be applied to the design of other HCI solutions for radiological applications that involve the capture of visual data and human descriptions of image findings.

## 1. Introduction

Conventional radiology reporting consists of radiologists dictating image findings from an image or series of images which are transcribed into a narrative text document. The radiology report serves as a means to convey diagnostic information to healthcare providers and to document a work product for billing purposes. The concept of structured reporting using standardized formats and terminology has shown promise recently as a means to improve the content and clarity of information and enable data mining [2], [3], [4]. Some structured reporting approaches utilizing menus and check boxes have proven to be too tedious and time-consuming for radiologists, thus leading to incomplete and inaccurate results compared to traditional narrative reporting [5]. The use of structured reporting today is generally limited to certain fields including breast imaging, cardiology, and gastroenterology where the anatomy and number of diseases are limited, thus allowing structured reports to be created with a few keyboard and/or mouse clicks[6]. To improve the efficiency of human interaction with these structured reporting systems and increase their clinical adoption, new data entry processes need to be created which incorporate the natural look-and-speak behavior of radiologist.

We have developed a novel structured reporting solution, called ViSion™, that captures key images during a radiologist's interpretive session along with verbal descriptions of the image findings, tags the key images with metadata, and generates an image-centric structured report [7]. To improve the efficiency of data entry, we have proposed the development of a human-computer interface (HCI) that captures eye-gaze and speech signals with an eye-tracking device. Prior literature on image perception has shown that regions-of-interest are associated with where a radiologist dwells upon for a long period of time [8], [9]. As a result, gaze signals have been employed in computer-assisted diagnosis and diagnostic systems to indicate regions-of-interest in either a single image or a series of images (i.e., image volume).

Our concept for improving the efficiency of data entry into the ViSion reporting system is to combine gaze signals with the cognitive information spoken by radiologists to create structured data. However, we recognize that radiology image interpretation and dictation is a complex task with many variables including the content of the image or volume being analyzed, interpretation task, screen content (e.g., hanging protocol), image modality, and human factors related to a particular radiologist (e.g., education, experience, age, visual acuity, alertness, etc.). Our research examines the feasibility of using eye-tracking technology and which factors need to be considered when constructing a gaze-speech HCI system for use in structured reporting.

## 2. Materials and Methods

The study protocol received Institutional Review Board approval. Written informed consent was obtained from the participating radiologists who performed the image interpretation and dictation experiments so that their gaze and speech signals could be studied. The radiological images used in this experiment were de-identified to comply with the Health Insurance Portability and Accountability Act (HIPAA) requirements.

Five board-certified radiologists at a major academic hospital were recruited to perform the experiments. The 5 radiologists consisted of: a) 3 fellows, 2 faculty, b) 4 male, 1 female, c) years of experience: (mean: 3.6, standard deviation: 4.2) Images from 3 modalities (CT, PET, Chest x-ray) were used with imaging parameters that are typically encountered in daily practice (e.g., spatial resolution, gray scale/color, contrast, and anatomical content). The experimental design was to give each radiology a particular set of image interpretation and dictation tasks for each set of images. The tasks involved a visual search for two distinct targeted findings in the same order for each modality and to speak about each finding. The image types and tasks were as follows:

a) Computed Tomography (CT) images of the liver (resolution 512x512): Identify and comment upon the largest liver metastasis followed by the abdominal aorta.
b) Position Emission Tomography (PET) of the whole body (resolution 512x512): Identify and comment upon the most significant metabolic abnormality followed by the degree of distention of the bladder.
c) Chest x-ray images (resolution 2020x2022): Identify and describe a lung lesion followed by the heart size.

An eye tracking device (Tobii T60 XL, Tobii Technology, Danderyd, Sweden) [10]was used to record eye gaze movements at 60 Hz, and a separate microphone was used to record the voice at 11025 Hz (Figure 1). A C# .NET software program was created to register each radiologist and calibrate the eye tracker to their gaze and vision. The 10 images for each modality were then shown in series to simulate a radiologist's daily workflow in a controlled environment. Figure 2 gives illustrates the interface shown to the radiologists during the experiments. Figure 3 represents the image types used for the experiments with scan paths for one of the participants.

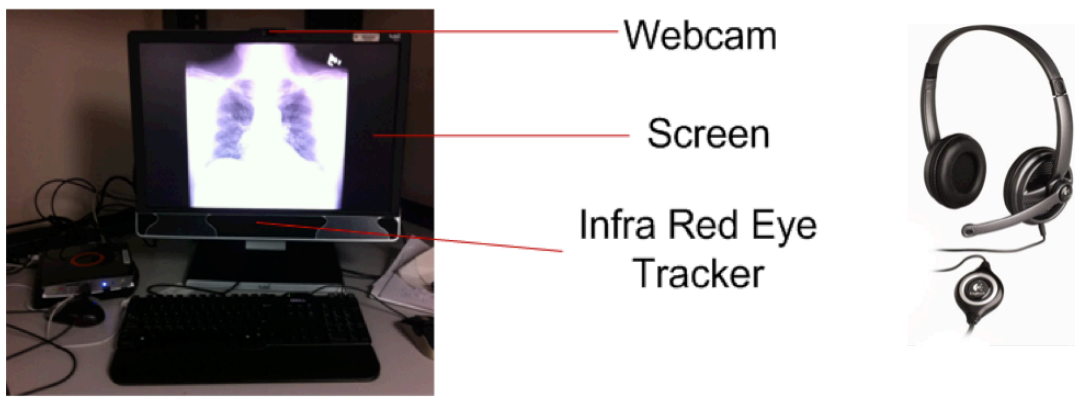

**Figure 1: Experimental setup and devices used. The eye-tracker on the left is the Tobii T60 XL that contains an embedded webcam and infrared tracker to monitor eye movements focused on the computer screen.**

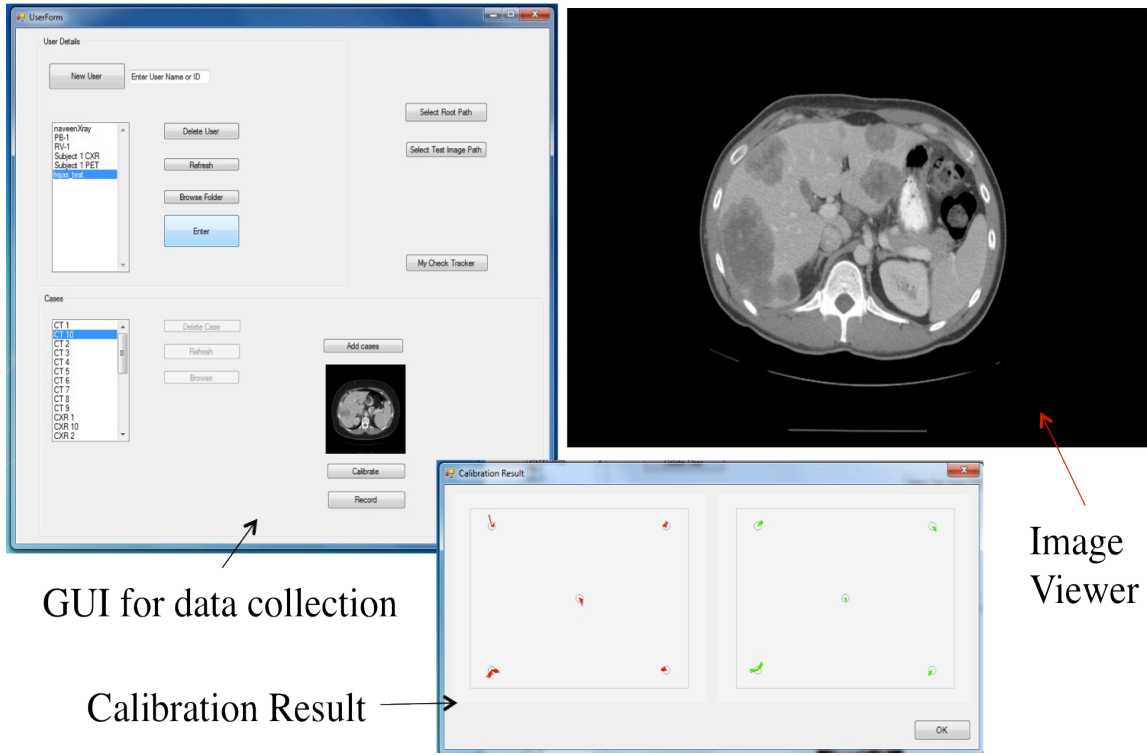

Figure 2: Graphical user interface (GUI) for data collection. Each radiologist must perform a calibration procedure with the eye tracker (bottom picture) in order to register eye movements with a certain accuracy. The image viewer shows each image in a full screen mode. The interface was programmed using a .NET platform.

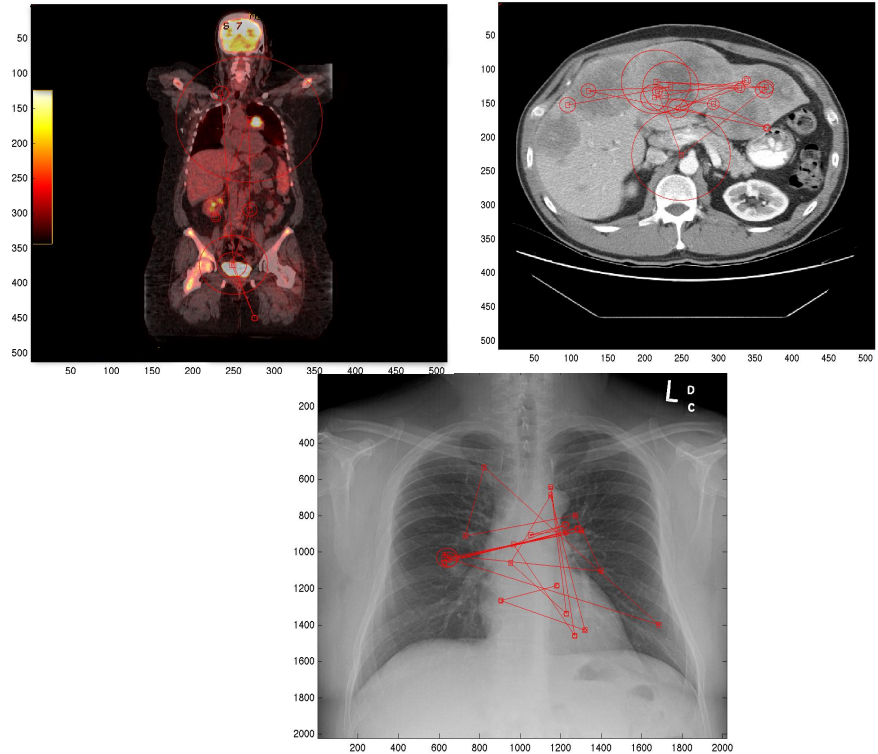

**Figure 3:** These figures are representative of the image types used in this experiment. The top image is a PET scan of the whole body (A), the middle is a CT scan of the liver (B), and the bottom image is a chest x-ray(C). The red overlays correspond to the fixations and dwells for a particular radiologist, and the radii of the circles correspond to dwell times.

### 2.1 Processing of Eye Gaze and Voice data

In order to analyze the temporal relationship between eye gaze recordings and subsequent voice recordings, we used an eye gaze–voice span parameter which is defined as the time difference between the onset of a fixation and the start of the corresponding speech. It is established in the literature on cognition that human beings always see a target before describing it verbally [11], [12]; hence the eye gaze–voice span is a defining parameter in the gaze–speech temporal relationship.

Dwell or fixation refers to the focus of gaze at a particular region-of-interest on the medical image. A visual angle of $2^o$ was assumed at the fovea. Fixations or dwells were computed from the eye tracking data by using this assumption, by the method highlighted in [13]. Figure 4 shows the overlay of raw eye gaze and dwell fixations on a representative PET image.

To identify the corresponding voice segment from the speech signal, manual annotation was performed on each of 30 speech signals (corresponding to 30 images) used for this experiment.

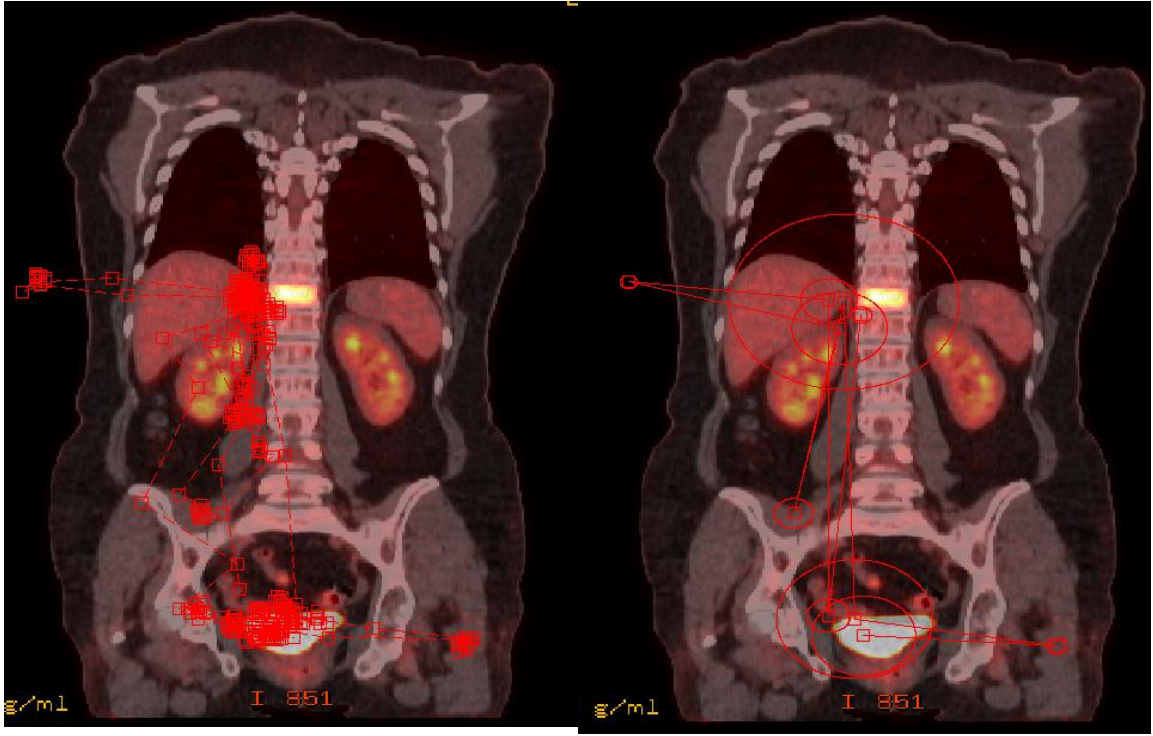

**Figure 4: The left side depicts raw eye gaze coordinates super imposed over the image. The right side depicts the fixation or dwells computed, superimposed over the same image. The time axis is collapsed for this illustration**

**2.2 Modeling of the gaze speech relationship**

We have considered the following key variables which could vary during an image interpretation and dictation session: a) image(s) being viewed, b) hanging protocol or viewing mode, and c) latent factors which are specific to the radiologist, such as alertness, education/ experience, and cognitive factors. The first two factors represent the screen content during the image interpretation session. In our experiments, we partitioned these factors into the following variables:

a) Image modality (CT/ PET/ Chest x-ray): The particulars of the images used from these modalities and the tasks assigned for each modality are presented in Section 2. This variable represents the screen content reflecting the modality, anatomy that was imaged, and image resolution.
b) Radiologist performing the dictation, which can be affected by education, experience, visual acuity, alertness, etc.
c) Target order for an interpretation task: the identification of targets is dependent. To elaborate, if $T_i$ corresponds to Target $i$, then the choice of $T_i$ depends on some subsequence of $\{T_1,...,T_{i-1}\}$. Consideration of this factor would provide insight into whether the eye gaze voice span varies significantly during a dictation across different targets, assuming all other factors are constant. The intra observer variability, i.e., variability within the same radiologist during an image interpretation diagnosis task is captured with this factor.

The eye gaze-voice span is modeled as a function of the radiologist, modality, and target order using a linear mixed effects model. The above factors are modeled as fixed effects. The identity of a particular image is referred to as the Image ID. The image ID is taken into account as a random effect in the model since the sample of images chosen for the experiments are a subset of a larger population of radiological images found in usual practice.

Mathematically, we model the eye gaze voice span ($y$) for a particular target in seconds as $y = f(rad,\text{mod},t)$, where $rad \in \{1,2,3,4,5\}$ is the radiologist, $\text{mod} \in \{CT,PET,XRay\}$ is the modality being viewed, and $t \in \{1,2\}$ is the target number in the dictation. Our goal is to fit this concept into a linear model. Each of the factors of interest, i.e., $rad,\text{mod},t$ is modeled as a fixed effect. Since the sample of images chosen for the experiments are a subset from the larger population of radiological images from usual practice, the image ID is modeled as a random effect. We have a total of 300 fixation-speech pairs to study the gaze speech relationship (5 radiologists x 3 modalities x 10 images/modality x 2 targets/image). Since we are adding a random intercept corresponding to image ID, we have a total of 30 groups in our model corresponding to each image.

Let $i$ be the group number and $j$ be the index of the data point in group $i$. We then have,

$$y_{ij} = \alpha_{i,0} + \beta_R rad_{ij} + \beta_M \text{mod}_{ij} + \beta_T t_{ij} + \hat{U}_{ij} \tag{2.2.1}$$

where $\alpha_{i,0}$ is the intercept corresponding to group $i$, $\hat{U}_{ij}$ is the error term per data point, and $\beta$s correspond to the coefficients of the model. Since the eye gaze voice span is a continuous quantity through a natural phenomenon, we assume that $\hat{U}_{ij} \in N(0,\sigma^2 \lambda_{ijj'})$ and $\text{cov}(\hat{U}_{ij},\hat{U}_{ij'}) = \sigma^2 \lambda_{ijj'}$ where $\lambda_{ijj'}$ are the covariances between errors in group $i$. Let us simply the model by assuming that the data points in group $i$ are i.i.d. Then the covariance matrix is diagonal and $\hat{U}_{ij} \in N(0,\sigma^2)$. This translates to assuming that the noise (error term) for each fixation speech pair is independent of the image or target order, which is a reasonable assumption to make.

Since we want to model the image ID as a random effect,
$$\alpha_{i,0} = \beta_{i,0} + u \tag{2.2.2}$$
where $u \sim N(0,\phi^2)$.

Combining equations (1) and (2), the following is our final linear model representing the gaze speech data,

$$y_{ij} = \beta_{i,0} + \beta_R rad_{ij} + \beta_M mod_{ij} + \beta_T t_{ij} + \hat{U}_{ij} \qquad (2.2.3)$$

The lme() package in R was used to estimate the regression coefficients, corresponding standard errors, and statistical significance from the model.

## 2.3 Visualization and Analysis of gaze probability maps

The purpose of this analysis is to calculate the probability of gaze at a particular coordinate $(x,y)$, given the image modality and the radiologist, in order to analyze and visualize the variation of gaze paths with these 2 factors. The probability of gaze is visualized as a gaze probability map, and its value at a particular coordinate point $(x,y)$ is a measure of the likelihood that the pixel $(x,y)$ was a gaze point.

Let $x_{M_1} \ldots x_{M_n}$ be images from modalities $M \in \{PET, CT, XRay\}$. Let $g_{M_1}, \ldots, g_{M_n}$ be the gaze maps corresponding to the images $x_{M_1} \ldots x_{M_n}$. Corresponding to each $x_{Mi}$ we have a gaze map $g_{Mi}$ defined as follows:

$$g_{M_i}(x,y) = \begin{cases} 1 & (x,y) \in \{GazeCoordinates\} \\ 0 & (x,y) \notin \{GazeCoordinates\} \end{cases} \qquad (2.3.1)$$

The content of these $n$ images comprise of the same anatomical information across different patients. Therefore, in order to use the eye gaze data from these images, we need to register both these images and the eye gaze data to a common coordinate system. This enables the possibility of superimposing gaze maps and images of a particular modality, in order to analyze the pattern of gaze of a radiologist for a particular image type (image modality with certain anatomical content).

In order to align the data such that the images $x_{M1}, x_{M2}, \ldots x_{Mn}$ match the alignment of $z$, the images are registered to the template $z$, using the algorithm described in [14]. The registration algorithm, which is based on maximizing mutual information, returns $A_{rot}$ which is a locally affine transformation that performs rotation/translation or warping on $x_{Mi}$ to produce $x'_{Mi}$, given by, $x'_{Mi} = A_{rot} x_{Mi}$. Similarly $g'_{Mi} = A_{rot} g_{Mi}$.

We assume a 2 degree visual angle at the fovea to account for gaze uncertainty. Using the parameters of the dictation environment, namely, size of the screen, distance of radiologist from the screen, we calculated the number of image pixels contained in the 2 degree diameter around each gaze point.

The transformed gaze maps are then convolved with a Gaussian Kernel with a width corresponding to 2 degrees in pixels.
Let $h$ be a Gaussian kernel with standard deviation $\sigma = k/2$, where $k$ is twice the number of image pixels contained in the 2 degree visual angle.

We then obtain the following,

$$g^{f}_{M_i} = g_{M_i} \star h$$

where $\star$ is the linear convolution operator.

The resulting maps are then normalized to produce a probability distribution function across pixels, per radiologist and per modality. The aggregate probability distribution conditioned on the radiologist and the image modality is, hence, calculated as follows, where $P_{map}$ is normalized to have a maximum value of 1.

$$P_{map} = \frac{1}{n} \sum_{i=1}^{n} g^{f}_{Mi}$$

We use these to visually study the variation of gaze paths across radiologists, for different settings in the dictation environment. A question that might arise here is that: Aren't the location of lesions and targets different across each of the 10 images of a particular modality? To answer this, each of the 5 radiologists looks at the same 20 targets over 10 images per modality. Therefore, the aggregated gaze probability map $P_{map}$ of a particular modality is marginalized over the factor of image target and lesion, and is solely a function of the radiologist.

## 3. Results

This section contains two analyses: a) gaze-speech relationship as a function of variables in the radiology reading environment, and b) visual analysis of gaze paths as a function of the radiologists.

### 3.1. Analysis of Gaze Speech Relationship

As described in Section 2.2, the gaze-speech relationship is parameterized by the eye gaze-voice span, which is the time difference between the onset of a fixation and start of corresponding speech. Our dataset consists of 5 radiologists, each performing readings on 30 images (of 3 image modalities), where each dictation per image was designed to have 2 targets. This leads to 300 total eye voice spans (data points). Since we have a reasonable sample size, we conclude that our results have both clinical and statistical significance.

From the linear model presented in Section 2.2, we test the following null hypotheses:
1. The eye gaze voice span does not vary significantly with radiologist.
2. The eye gaze voice span does not vary significantly with image type.
3. The eye gaze voice span does not vary significantly with the target order

Our results indicate that there is no statistically significant evidence that the eye gaze-speech temporal relationship, parameterized by the eye gaze-voice span, depends on the image modality or the target order within a radiology dictation. However, the individual radiologist is a statistically significant factor affecting the eye gaze-voice span (Table 1).

**Table 1: Modeling of Gaze Speech Relationship parameterized by eye voice span as a function of Image Modality, Radiologist and Target Order: Linear Mixed-Effects Model analysis**

| Factor | p value |
| --- | --- |
| Modality | 0.909 |
| Radiologist | <0.0001 |
| Target order | 0.142 |

The above findings indicate that, among the factors taken into account, the individual radiologist is key in designing systems to automate structured reporting using a gaze – speech human computer interface. Hence, such systems would need to be customized per user or radiologist, and do not depend on the radiologist interpretation task, anatomy that was imaged, image modality or resolution. We can also conclude with good statistical significance, that the intra observer variability, i.e., variability within the same radiologist during an image interpretation diagnosis task, captured by the target order variable, is also not a factor of variation for such a potential system. Our statistical analysis indicates that such a eye gaze – speech HCI solution needs to be customized per user or radiologist.

The variables considered in our analysis, i.e., a) radiologist, b) image modality, c) target order, embeds the following factors in a real life dictation situation: a) radiologist dependent: education, training, cognitive process, skill, b) image modality dependent: image resolution, anatomy being imaged, modality, and c) target order: intra radiologist variability. However, we have not included factors such as ambient lighting, noise, temperature, radiologist alertness, etc.

### 3.2 Analysis of eye gaze paths

We have also visualized the aggregate gaze probability maps (calculated by the method presented in section 2.3), per radiologist and per modality. This visualization lends some more intuition to the statistical findings above which conclude that radiologist dependent factors are the most key to consider when designing an eye gaze and speech based HCI solution for structured reporting.

The probability maps are aggregated per radiologist and modality. In this contrived radiology reading experiment, each of the imaging modalities has 1 fixed target, which is constant for all images and one search target, which varies per image. As explained in Section 2.3, the aggregate gaze probability map per modality and radiologist marginalizes over the targets. To elaborate, all of the 5 radiologists viewed the same targets over 10 images in each modality. Across images, there was one target that was fixed: abdominal aorta for CT, bladder for PET, and heart for x-ray. Therefore, intuitively, we should expect that all the 5 radiologists spent some period of time looking at these fixed targets after performing the interpretation task on the variable target: largest liver metastasis for CT, most significant abnormality for PET, and a lung lesion of choice for x-ray. Since, a larger pixel value (brighter pixel) denotes a higher probability of gaze, we should see a concentration of gaze around the fixed targets. Figure 5, 6, and 7 show probability maps that visually depict the probability of a particular region-of-interest in an image to be gazed upon by a radiologist for all 3 imaging modalities considered.

Figure 5 depicts the gaze probability maps for the 5 radiologists for the Chest x-ray modality. We see that while one radiologist (right top corner) is focused on certain areas of the image during the visual search, another radiologist (middle bottom) scans more areas of the image on an average, given the same visual task. Due to the nature of the visual task in this experiment, all the radiologists show a high gaze probability around the heart.

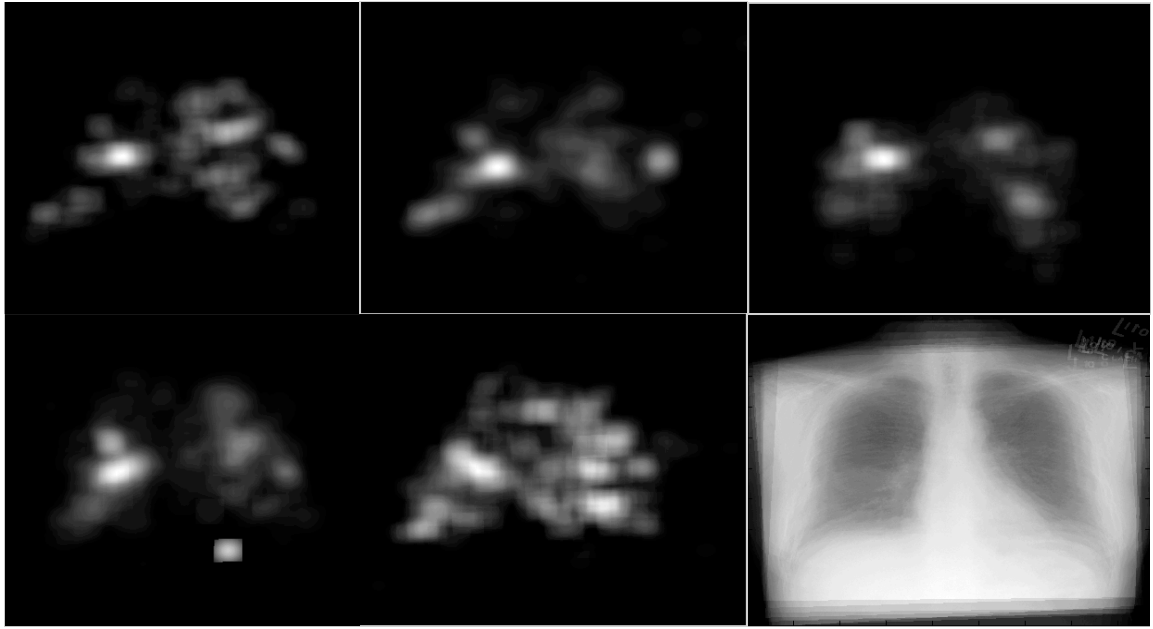

Figure 5: Gaze probability maps for the 5 radiologists for the chest x-ray modality. The image on the bottom right is the average image of all images shown to the radiologists, registered to a common coordinate system. Although this image doesn't contain details of the individual lesions, it provides a visualization of the anatomy. The pattern of gaze varies for the different radiologists, although all of them have a high gaze probability around the heart region. (White: high probability, Black: Low probability)

We see a similar trend in the gaze probability maps for the CT modality (Figure 6). Here, the variation in gaze patterns across radiologists is primarily in the dwell time associated with the search targets. The radiologist on the top left corner appears to perform a quicker search across the image, compared to the radiologist on the bottom left. All the radiologists show a high gaze probability around the liver, which contained a lesion and was the nature of the visual task, similar to the heart in the chest x-ray experiments.

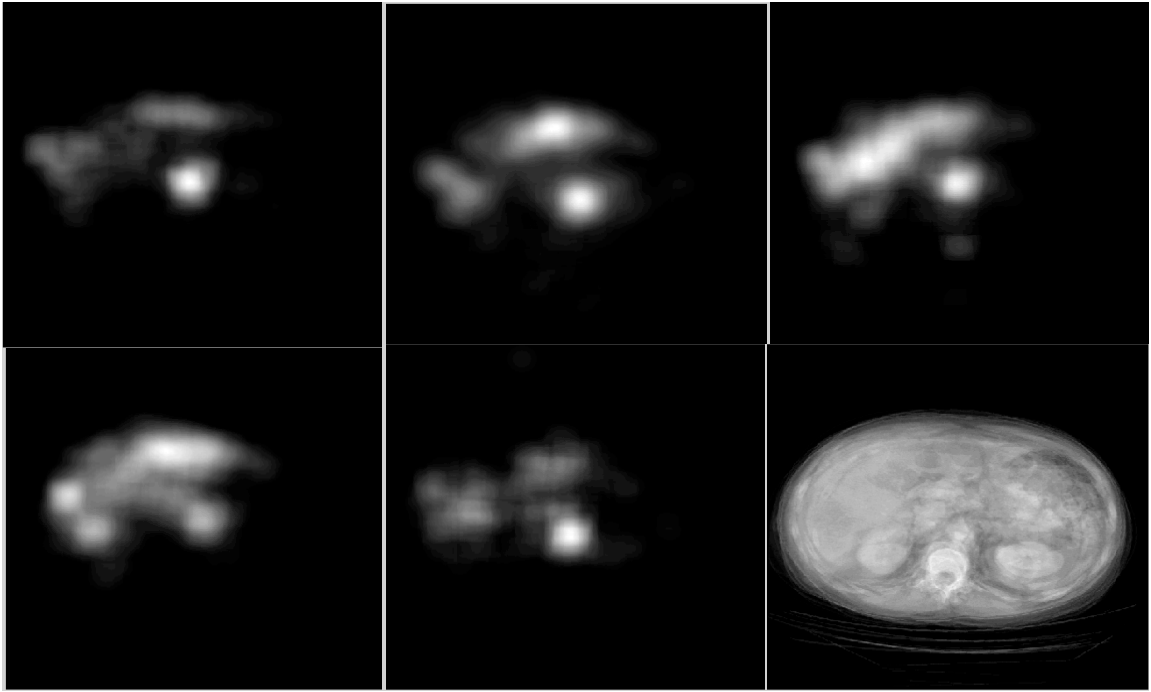

Figure 6: Gaze probability maps for the 5 radiologists for the CT modality. The image on the bottom right is the average image of all images shown to the radiologists, registered to a common coordinate system. The pattern of gaze is different across radiologists, although all of them have a high gaze probability around the liver region.

For the PET images (Figure 7), we see a high probability region around the bladder as expected. Contrary to the other modalities, the non-fixed task, i.e., identification of the most significant metabolic activity seems to have required a whole body scan for all of the radiologists. Since all the radiologists were looking across the whole body to identify different metabolic activities across different images, the aggregate probability maps have high probability pixels across the human body. However, even in this case, the number of high probability regions (white in the aggregate probability map) is not identical per radiologist.

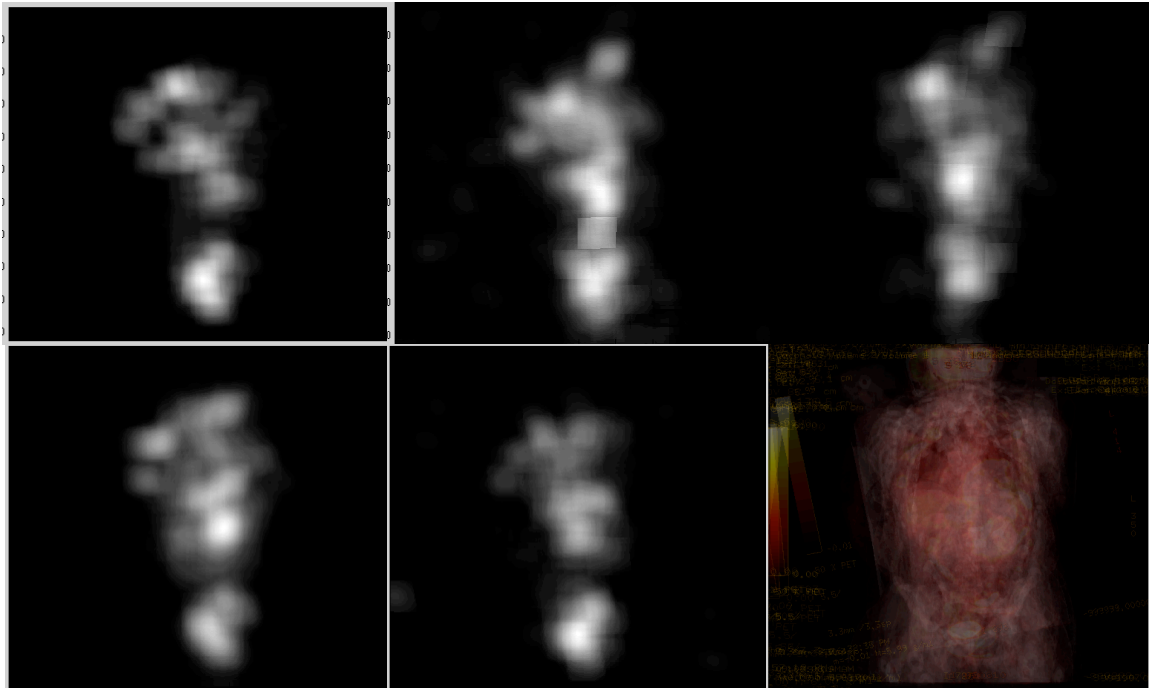

Figure 7: Gaze probability maps for the 5 radiologists for the PET modality. The image on the bottom right is the average image of all images, registered to a common coordinate system.

Our analysis of eye gaze paths through aggregate probability maps provides better intuition and supports our earlier conclusion that gaze paths vary significantly among radiologists, and hence an HCI solution for structured reporting would need to be customized for each radiologist.

It is important to recognize that these findings were observed in a controlled experimental environment that attempted to simulate a natural workflow environment. Eye movements and gaze paths are complex and could be more varied in a real working environment. The dictation tasks were designed in such a manner that the radiologist was instructed to look at a target before speaking about it to control the experiment. In reality, there is a possibility that a radiologist might identify a second target before concluding speaking about the 1st target. This would add extra complexity to both the gaze paths and the gaze-speech relationship, thus making the radiologist an even more important factor in customizing/designing a gaze-speech HCI, since image interpretation for diagnosis is a highly cognitive task.

## 4. Discussion

In this study, we evaluated the factors related to a potential eye gaze–speech based human computer interaction solution for automating structured reporting. We know that the eye gaze path is representative of the visual search for diagnostic findings by a radiologist [15], [16]. The eye gaze of a radiologist is directed towards the targets specified by the search task and are congruent with the visual task in diagnostic imaging [17], [18]. Therefore, the eye gaze signal, when fused with the voice signal during image interpretation, contains the essential diagnostic information, including findings (defined by image coordinates) and diagnoses (voice descriptions) at different points of time, needed to construct a radiology structured report in accordance with our ViSion process.

We designed a set of dictation experiments which were controlled so we could measure various factors in a real-life radiology image interpretation and dictation environment. The experiments were conducted with 5 radiologists. In each pair of eye gaze and speech signal, a radiologist always looks at a target before verbally describing it. This is consistent with prior literature in the field of human communication and HCI [12], [19].

The statistical analysis for this study was performed on the eye gaze-voice spans. Since we have 5 radiologists, 10 images for each of 3 modalities, each image containing 2 targets, thus resulting in 300 pairs of eye gaze – voice spans, there is good statistical significance of our findings. On fitting a linear mixed effects model to the eye gaze – voice spans, we concluded that the key factors to be taken into account when designing such an HCI system would be radiologist dependent, i.e., the system would need to be customized to the profile of the radiologist, since each radiologist has a distinct cognitive process by which he or she performs image interpretation and dictation tasks. Factors such as a radiologist's education, skill, weariness, are additional variables that might affect the cognitive process but were not measured in our experiment. It is not clear whether the style or the cognitive process of the radiologist would change with time and age, but if so, then the potential HCI system may need to be re-calibrated and re-customized throughout the career of a radiologist; however, this aspect was not investigated in this study. We also concluded that intra-radiologist variability, captured by the 'target order' variable is not a significant variable. In addition, the anatomy being imaged, image resolution, and modality do not need separate customization for an eye gaze – speech based HCI system.

Given the eye gaze data we also calculated aggregate probability maps for each radiologist. Our analysis of these probability maps confirms the statistical conclusions and provides visual intuition on how search patterns vary among radiologists given the same set of images and targeted findings.

The next step would be to use the gaze and speech signals together to identify diagnostic regions-of-interest from the gaze signals and corresponding speech content to incorporate key image findings into a structured report. Prior literature on visual search strategies of radiologists developed from analyzing gaze data indicates that regardless of the variability of gaze patterns among radiologists, they follow a visual strategy. The visual strategy starts with a holistic look at the entire image or volume followed by a closer detailed look for a given interpretation task [16]–[18].

## 4.1 Limitations

In the analysis of our experimental design, we attempted to capture certain variables in a radiologist's reading environment, which were a function of the radiologist, image modality, and target order. These variables comprised factors such as image resolution, anatomy being imaged, intra-radiologist variability and other factors related to the cognitive process of a radiologist. However, considering that real-life image interpretation and dictation is complex, there could be case- and image-dependent factors which were not considered in our experiments. For example, the images that we used were static 2D images and not a series of images through which a radiologist scrolls in a complete CT or MRI examination. We also did not consider ambient factors in the dictation environment such as lighting, noise, distractions, etc.

This analysis was performed using a simulated and controlled reading environment. During this experiment, the radiologist always looked at a specific target and subsequently spoke about that particular finding. In a real-life scenario, a radiologist might look at a second target while continuing to speak about the first target, thus increasing the complexity of the temporal gaze-speech relationship. We hypothesize that even in this complex scenario, the eye gaze-speech dynamics and gaze paths will differ significantly between radiologists, although we cannot make claims with certainty about intra-radiologist variability without further study.

Previous studies have indicated that the visual search strategy varies between experts and novice readers [12], [19]. With only 5 radiologists comprising faculty and fellows, we could not verify this claim.

## 4.2 Practical Implications

This knowledge on gaze patterns and gaze-speech temporal relationships may be used to create data entry processes for radiology structured reporting. An HCI-based solution should mimic the efficiency of a conventional transcription-based narrative reporting while simultaneously creating structured data and eliminate the cumbersome manual interaction required by some structured reporting systems. This could lead to more widespread adoption of structured reporting since excessive human interaction (e.g., mouse clicks) required to create some types of structured reports has been identified as a significant hurdle [2], [20], [21].

The findings from this research could be extended beyond structured reporting to other radiology applications that utilize an eye gaze-speech HCI.

## Conclusion

This research established that the most important variable to be considered when designing a human-computer interaction solution based on eye-gaze and speech for structured reporting are those that are a function of the actual radiologist. Factors related to screen content and intra-radiologist variance are independent. This was concluded by statistical modeling of the eye gaze–voice spans with good statistical significance, and further intuition was provided by viewing visualizations of probability gaze maps.

Since we have established that factors which depend solely on the radiologist are the most important when designing an HCI, we recommend that future research in the domains of structured reporting, visual search in radiology, perception of medical images, and even machine learning solutions be focused on training and conditioning the radiologist to better use the HCI.